\documentstyle[12pt]{article}
\begin{document}
\title{The Universe of Chaos and Quanta}
\author{B.G. Sidharth\\
Centre for Applicable Mathematics \& Computer Sciences\\
B.M. Birla Science Centre, Adarsh Nagar, Hyderabad - 500 063 (India)\\
Email:birlasc@hd1.vsnl.net.in; birlard@ap.nic.in}
\date{}
\maketitle
\begin{abstract}
In recent papers it was shown that stochastic processes in the universe
as a whole lead to discrete space time at Compton scales as also non-
relativistic Quantum Mechanics. In this paper,
we deduce the Dirac equation and thence a
unified formulation of quarks and leptons. In the process several puzzling
empirical results and coincidences are shown to be a consequence of the
theory. These include the discreteness of the charge, handedness of quarks, their fractional charge,
confinement and masses and the handedness of neutrinos, the so called
accidental relation that the classical Kerr-Newman metric describes, the
field of an electron including the purely quantum mechanical gyromagnetic
ratio $g=2$, as also the many large number coincidences made famous by
Dirac and Weinberg's mysterious empirical formula that relates the pion
mass to the Hubble Constant. A cosmology based on fluctuations related to
the above stochastic space-time discretization, consistent with latest observations
is also seen to follow.
\end{abstract}
\section{Introduction}
In previous communications\cite{r1,r2,r3} an attempt was made to
give a stochastic formulation as the underpinning to physical laws. The
starting point was the well known Brownian random walk relation
\begin{equation}
R = l \sqrt{N}\label{e1}
\end{equation}
which is true for the universe as a whole, $R$ being its radius, $N$ being
the number of elementary particles, typically pions in the universe and
$l$ the pion Compton wavelength. It was argued that (\ref{e1}) would be
the starting point as a rationale for quantum mechanics itself. We will
now carry the argument one step further and show that, on this basis we can
deduce the two basic equations of quantum mechanics, namely the Dirac and
Schrodinger equations and thence deduce a model for all the fundamental
particles namely the quarks and the leptons. This model will bring out all
peculiar characteristics of these basic particles, including features like
fractional charges, handedness, confinement and so on. Hitherto these
features have been purely empirical without a theoretical rationale.
At the same time, it will be
pointed out that in the same vein one could get a cosmological description,
consistent with data and observation.
\section{An Underpinning of Quantum Theory}
We first observe that, using the well known fact that $R = cT$, where $T$
is the age of the universe, the following relation can be obtained from
(\ref{e1}):
\begin{equation}
T = \sqrt{N} \tau\label{e2}
\end{equation}
where $\tau = l/c$ is the pion Compton time.\\
Furthermore, as there are $N$ particles typically pions in the universe,
it should follow that
\begin{equation}
M = Nm\label{e3}
\end{equation}
where $M$ is the mass of the universe and $m$ the pion mass. This is indeed
so.\\
We next observe that equations (\ref{e1}) and (\ref{e2}) indicate that the
Compton scale is a fundamental unit of space time. Infact it was shown
\cite{r2} that this quantized space time could be considered to be more
fundamental than Planck's quantized energy - the latter can be shown to
follow from the former. It may be mentioned that the concept of discrete
space time has a long history and has been studied by several authors for
example Snyder, Schild, Kardyshevsky, Bombelli, Wolfe and several others
\cite{r4}-\cite{r8} As T.D. Lee observes\cite{r9}, the space-time
continuum is an approximation. Infact as pointed out\cite{r2,r10} such a quantized space time description, following from (\ref{e1})
and (\ref{e2}) provides a rationale for a maximal velocity and Special
Relativity itself. Snyder\cite{r4} pointed out that discrete space time is
compatible with Special Relativity and went on to show
$$[x, y] = (\imath a^2 / \hbar)L_{z,}  [t, x] = (\imath a^2 / \hbar c)M_{x,}$$
\begin{equation}
[y, z] = (\imath a^2 / \hbar) L_{x,} [t, y] = (\imath a^2 / \hbar c)M_{y,}\label{e4}
\end{equation}
$$[z, x] = (\imath a^2 / \hbar) L_{y,} [t, z] = (\imath a^2 / \hbar c)M_{z,}$$
where $a$ is the minimum natural unit and $L_x, M_x$ etc. have their usual significance.\\
Similarly it was also deduced that
$$[x, p_x] = \imath \hbar [1+(a/\hbar)^2 p^2_x];$$
$$[t, p_t] = \imath \hbar [1-(a/ \hbar c)^2 p^2_t];$$
\begin{equation}
[x, p_y] = [y, p_x] = \imath \hbar (a/ \hbar)^2 p_xp_y ;\label{e5}
\end{equation}
$$[x, p_t] = c^2[p_{x,} t] = \imath \hbar (a/ \hbar)^2 p_xp_t ;\mbox{etc}.$$
where $p^{\mu}$ denotes the four momentum.\\
While equations (\ref{e4}) bring out the non commutative nature of space time
geometry, equations (\ref{e5}) show a departure from the usual quantum
mechanical commutation relations. However as $a \to 0,$ we recover the
usual theory, as expected.\\
Using now the fact that the minimum natural unit $a$ is the Compton wavelength,
$l = \hbar/mc$, and the fact that at these extreme scales $|\vec p | = mc,$
it follows from (\ref{e5})
\begin{equation}
[x, p_x] = 2 \imath \hbar\label{e6}
\end{equation}
and similar equations.\\
The right hand side of (\ref{e6}) is double the usual value - as if the spin has been halved or
the coordinate doubled. This is the typical double connectivity of spin half.\\
This is suggestive of the surprising fact that Quantum Mechanical spin would
arise from quantized space time. This can be confirmed as follows: We
consider a linear transformation of the wave function,
\begin{equation}
|\psi' > = U(R) |\psi >\label{e7}
\end{equation}
If the transformation $R$ is a simple infinitesimal  coordinate shift in Minkowski space
we will get, from (\ref{e7}),  as is well known\cite{r11,r12},
\begin{equation}
\psi' (x_j) = [1 + \imath \epsilon (\imath \epsilon_{ljk} x_k \frac{\partial}
{\partial x_j}) + 0 (\epsilon^2)] \psi (x_j)\label{e8}
\end{equation}
We next consider the commutation relations (\ref{e4}), taking $a$ to be the
Compton wavelength. We can easily verify that the choice
\begin{equation}
t =  \left(\begin{array}{l}
        1 \quad 0 \\ 0 \quad -1 \\
        \end{array}\right), \vec x =
   \ \  \left(\begin{array}{l}
         0 \quad \vec \sigma \\ \vec \sigma \quad 0 \\
       \end{array}\right)\label{e9}
\end{equation}
provides a representation for the coordinates
$x$ and $t$ apart from any scalar factors. We can immediately recognize that
equation (\ref{e9}) is a representation
of the Dirac $\gamma$ matrices. Substitution of (\ref{e9}) in (\ref{e8}) now
leads to the Dirac equation,
\begin{equation}
(\gamma^\mu p_\mu - mc^2) \psi = 0\label{e10}
\end{equation}
because
$$E\psi = \frac{1}{\epsilon} \{ \psi' (x_j)-\psi
(x_j)\}, E=mc^2,$$
where $\epsilon = \tau$ (cf.ref.\cite{r8}).
Thus we have shown that given minimum space time intervals of the order of
the Compton wavelength and time, a simple coordinate transformation leads to the Dirac
equation. (Alternatively, we could have considered the operator
$\exp (-\imath\ \epsilon^\mu p_\mu)$.)\\
It must be mentioned that the usual derivation of the Dirac equation itself
is heuristic, in the sense that the square root operator is not very
transparent, while at the same time the emergence of a four component spinor
instead of a two component spinor and the negative energy solutions and
non Hermitian position operators are puzzling, as also the standard interpretation
that on averaging over Compton scales, these Zitterbewegung effects and non
hermitian operators disappear. All this now becomes transparent - physics begins
outside these minimum quantized space time Compton scales so that we obtain
a physical interpretation only after averaging over these intervals and we
encounter unphysical effects at smaller scales\cite{r13,r14,r15,r9}).
That is, a space time continuum is an approximation that breaks down within
the Compton scales.\\
It is ofcourse well known that in the non relativistic limit $v/c < < 1$, the
Dirac equation goes over to the Schrodinger equation. Infact it was shown
that the above Compton scale cut off\cite{r2} explains the otherwise adhoc and
therefore unsatisfactory features of Nelson's stochastic derivation of the
Schrodinger equation\cite{r16}.\\
All this is a reconfirmation of the fact pointed out earlier that space time
quantization at the Compton scale provides an underpinning for quantum
mechanics.
\section{Quarks and Leptons}
At the Compton scale, as is known, the negative energy two spinor $\chi$ of
the full four rowed Dirac spinor $(\theta_\chi)$, (where
$\theta$ denotes the positive energy two spinor), begins to
dominate. Further under reflections, while $\theta$ goes to $\theta, \chi$
behaves like a psuedo-spinor\cite{r12},
$$\chi \to - \chi$$
Hence the operator$\frac{\partial}{\partial x^\mu}$ acting on $\chi$, a
density of weight $N = 1,$ has the following behaviour\cite{r13},
\begin{equation}
\frac{\partial \chi}{\partial x^\mu} \to \frac{1}{\hbar} [\hbar \frac{\partial}
{\partial x^\mu} - NA^\mu] \chi\label{e11}
\end{equation}
where,
\begin{equation}
A^\mu = \hbar \Gamma_\sigma^{\mu \sigma} = \hbar \frac{\partial}{\partial x^\mu}
log (\sqrt{|g|})\label{e12}
\end{equation}
$\Gamma$'s being the usual Christoffel symbols.\\
We can easily identify $NA^\mu$ in (\ref{e11}) with the electromagnetic
four potential. (That $A^\mu$ is a four gradient poses no problem\cite{r13}.
The fact that $N = 1$ explains why the charge is discrete). We can also
immediately see the emergence of the metric tensor and the resulting
potential.\\
On the face of it (\ref{e12}) is mathematically identical to Weyl's Unification
formulation of electromagnetism and gravitation\cite{r17}. However there is
a very important difference - (\ref{e11}) and (\ref{e12}) arise from the
purely quantum mechanical behaviour of the Dirac spinor (cf. also\cite{r18})
of the equation (\ref{e10}).\\
We now use the fact that the metric tensor $g_{\mu \nu}$ resulting from (\ref{e12})
satisfies an inhomogenous Poisson equation\cite{r17,r19},
\begin{equation}
\nabla^2 g_{\mu \nu} = G \rho u_\mu u_\nu\label{e13}
\end{equation}
It immediately follows that
\begin{equation}
g_{\mu \nu} = G \int \frac{\rho u_\mu u_\nu}{|\vec r - \vec r'|}
d^3\vec r\label{e14}
\end{equation}
where now we require the volume of integration to be the Compton volume.
Alternatively we observe that as is known\cite{r20,r21}, for the Dirac
equation (\ref{e10}) we can get the Hamilton-Jacobi formulation and thence
equation (\ref{e13}) when the energy term as in (\ref{e11}) and (\ref{e12}) is introduced
in (\ref{e10}).\\
In any case (\ref{e14}) can be recognized as the linearized general
relativity equation\cite{r19}. From (\ref{e14}) we can get the usual gravitational
potential for $r \equiv | \vec r - \vec r' | > >$ the Compton wavelength,
\begin{equation}
g_{00} = - \frac{Gm}{r}\label{e15}
\end{equation}
As shown elsewhere\cite{r13,r14,r18,r22},
given the linearized equation of General Relativity,
(\ref{e14}) was the starting point of a geometrized formulation of Fermions
leading to the Kerr-Newman metric and which explains the remarkable and supposedly
coincidental fact that the Kerr-Newman metric describes the field of an
electron including the anomalous gyro magnetic ratio $g = 2$.\\
In this connection two points need to be noted. Firstly an electron is
indeed a Kerr-Newman Black Hole except for a naked singularity. Secondly
in the derivation of the Kerr-Newman metric, an imaginary shift of origin
is used which as Newman confessed is as yet inexplicable\cite{r23}. Both
these have an immediate parallel in the Zitterbewegung and non Hermitian
position operators of the Quantum Mechanical electron. Once the minimum
Compton scale as above is recognized, these inexplicable features disappear.\\
All this was also
shown to lead to a unified description of electromagnetism, gravitation
and strong interactions\cite{r18,r22}.\\
We now show how a unified description of quarks and
leptons can be obtained from (\ref{e14}).\\
From (\ref{e12}) and (\ref{e14}) we get
\begin{equation}
A_0 = G \hbar \int \frac{\partial}{\partial t} \frac{(\rho u_\mu u_\nu )}
{|\vec r - \vec r'|} d^3r \approx \frac{ee'}{r}\label{e16}
\end{equation}
for $|\vec r - \vec r'| > >$ the Compton wavelength where $e' = e$ is the
test charge.\\
Further, from (\ref{e16}), as in the discrete case,
$d \rho u_\mu u_\nu = \Delta \rho c^2 = mc^2$ and $dt = \hbar/
mc^2$, we get
$$A_0 = \frac{e^2}{r} \sim G \frac{\hbar}{r} \frac{(mc^2)^2}{\hbar}$$
or
\begin{equation}
\frac{e^2}{Gm^2} \sim 10^{40}\label{e17}
\end{equation}
(\ref{e17}) is the well known but hitherto purely empirical relation expressing
the ratio of the gravitational and electromagnetic strengths. Here we have
deduced (\ref{e17}) from theory.\\
It was pointed out\cite{r18} that starting from
(\ref{e14}) one could obtain the Kerr-Newman metric which describes the field
of the electron, and via the equations (\ref{e11}), (\ref{e12}), (\ref{e14}),
(\ref{e15}), (\ref{e16}) and (\ref{e17}) we have affected the unified
description of electromagnetism and gravitation.\\
If in (\ref{e14}) we consider distances of the order of the Compton wavelength,
it was shown that we will get instead of (\ref{e15}), a QCD type potential
\begin{eqnarray}
4 \quad \eta^{\mu v} \int \frac{T_{\mu \nu} (t,\vec x')}{|\vec x - \vec x' |} d^3 x' +
(\mbox terms \quad independent \quad of \quad \vec x), \nonumber \\
+ 2 \quad \eta^{\mu v} \int \frac{d^2}{dt^2} T_{\mu \nu} (t,\vec x')\cdot |\vec x - \vec x' |
d^3 x' + 0 (| \vec x - \vec x' |^2) \propto - \frac{\propto}{r} + \beta r\label{e18}
\end{eqnarray}
where $T_{\mu \nu} \equiv \rho u_\mu u_\nu$. Equation (\ref{e18})
can lead to a reconciliation
of electromagnetism and strong interactions\cite{r22}. For this we need to
obtain a formulation for quarks from the above considerations. This
is what we will do now.\\
We first use the fact that the doubleconnectivity or spin half of the electron
leads naturally to three dimensional space\cite{r19, r24}, that is at asymptotic
distances far outside the Compton wavelength. However this doubleconnectivity
or spinorial behaviour breaks down at Compton scales and we need to consider low
dimensionality namely two and one dimensions. Using the well known fact that
each of the $T_{\imath j}$ in (\ref{e16}) is given by $\frac{1}{3} \epsilon$
\cite{r25}, $\epsilon$ being the energy density, it follows immediately
from (\ref{e16}) that the charge would be $\frac{2}{3} e$ or $\frac{1}{3}e$
in two or one dimensions, exactly as for quarks. At the same time as we are
now at the Compton scale, automatically these fractionally charged particles
are confined. This indeed is expressed by the confining part of the QCD
potential (\ref{e18}). Further, at the Compton scale, as noted earlier\cite{r12}
we encounter predominantly the negative energy components of the Dirac spinor
with, opposite parity so that these quarks would show neutrino type handedness,
which indeed is true.\\
Thus at one stroke, all the peculiar empirical characteristics of the quarks
for which as Salam had noted\cite{r26}, there was no theoretical rationale,
can now be deduced from theory. We can even get the correct order of
magnitude estimate for the quark masses. For this we rewrite (\ref{e18}) by
multiplying the expression by $\frac{1}{m}$ (owing to the factor $\frac{\hbar^2}
{2m}$), and at the same time go over to units $c = \hbar = 1$. This will
facilitate comparison with standard literature\cite{r27}. (\ref{e18}) then
becomes
\begin{equation}
\frac{4}{m} \quad \eta^{\mu v} \int \frac{T_{\mu v}}{r} d^3 x + 2m \quad
\eta^{\mu v} \int T_{\mu v} r d^3x \equiv
-\frac{\propto}{r} + \beta r\label{e19}
\end{equation}
where now, $\propto \sim O(\ref{e1})$ and $\beta \sim O(m^2),$
where $m$ is the mass of the quark in agreement with standard literature.\\
We now observe that at the Compton scale, for the fractionally charged quarks
$e^2$ in (\ref{e16}) goes over to $\frac{e^2}{9} \sim 10^{-3}$ in the units
being considered. So (\ref{e16}) can be written as
\begin{equation}
\frac{1}{r} \approx 10^3  G \hbar \int \frac{\partial}{\partial t}
\frac{\rho u_\mu u_\nu}{|\vec r - \vec r'|} d^3 r\label{e20}
\end{equation}
Comparing (\ref{e20}) for the quarks with the Coulumb term of the QCD
potential (\ref{e19}) and the electrostatic potential for the electron given
in (\ref{e16}), we can immediately see that the quark mass would be $\sim
10^3 \times$ the electron mass, which ofcourse is in order.\\
The above description is equally valid for neutrinos if we remember that they
have vanishingly small mass. So their Compton wavelength is very large and by
the same argument as above, we encounter predominantly the negative energy
components of the Dirac spinor which have opposite parity, that is the
neutrinos display handedness.\\
This brings us to the question of weak interactions. So far we have dealt with
interactions which in the conventional language, are mediated by mass less
intermediary Bosons.\\
We will now argue that a weak electromagnetic interaction with a massive
intermediary $m_w \sim 100m_p$, where $m_p$ is the proton mass, is entirely
consistent with the foregoing considerations.\\
In the early days of the electro weak theory\cite{r28} it was already
realized that this was indeed the case for the weak coupling constant roughly
set equal to the electromagnetic coupling constant:
\begin{equation}
G_w \equiv g^2/m^2_w \approx \frac{10^{-5}}{m^2_p} gm^{-2}\label{e21}
\end{equation}
where $G_w$ is the Fermi local weak coupling constant.\\
(\ref{e21}) can be deduced from (\ref{e17}), where it should now be borne
in mind that (\ref{e17}) is not an adhoc relation, but rather a deduction
from the theory. Infact with $g^2 \approx 10^{-1}$, we get from (\ref{e17})
and (\ref{e21}) $G_w \approx 10^{43}$ as required. We will come back to this
point again in section 5.
\section{Cosmological Considerations}
We now come to the field of Cosmology and will consider the fluctuational
creation of particles, which will be seen to be closely linked with the
stochastic minimum space time intervals. Our starting point is the fact that
as is well known there is a background Zero Point Field (ZPF), which
as noted by Vigier and others indeed
has been considered to be what drives fluctuations\cite{r29,r30,r31}. It is
also known that the energy of the fluctuations of for example the magnetic
field in a region of length $l$ is given by\cite{r19}
$$B^2 \sim \frac{\hbar c}{l^4}$$
where $\vec B$ is the magnetic field strength.\\
Taking now $l$ to be the Compton wavelength, for a volume $\sim l^3$, the
energy of the resultant particle is given by
$$\mbox{Energy}\quad \sim \frac{\hbar c}{l} = mc^2,$$
which ofcourse is correct. In other words the entire energy of an elementary
particle of mass $m$ is generated by the fluctuations alone\cite{r14}.
We now show that all this leads to a perfectly consistent cosmology.\\
As in the usual theory we equate the Newtonian gravitational
potential energy (which is dominant on a large scale) of the pion in a three dimensional isotropic
sphere of pions of radius $R$, the radius of the universe, with the rest
energy of the pion, to get the well known relation
\begin{equation}
R = \frac{GM}{c^2}\label{e22}
\end{equation}
where $M$ can be obtained from (\ref{e3}).\\
We now use the fact that the fluctuation in the particle number is of the
order $\sqrt{N}$\cite{r14,r32,r33}, while a typical time interval of uncertainity
is $\sim \hbar/mc^2$ as seen above. (That is particles induce
more particles by fluctuations).\\
It must be mentioned that this is somewhat in the spirit of the work of Prigogine
and his coworkers\cite{r34,r35}: Heisenberg's Uncertainity gives
rise to production of energy over short intervals of time, leading to a one
way creation of particles. In the words of Prigogine\cite{r36}, "...in this perspective,
time is eternal... has neither a beginning nor an end. We come therefore to
a position which unifies elements contained in the two traditional uses of
cosmology: The Steady State Theory and the standard Big Bang
approach."\\
In this light we combine the usual fluctuation in particle number with the
minimum uncertainity in time to consider the creation of particles, from the
background ZPF.\\
This leads to the relation\cite{r14},
$$\dot N \equiv \frac{dN}{dt} = \sqrt{N} \frac{mc^2}{\hbar}$$
whence an integration between
$t = 0$ and $t = T$, the present age of the universe, gives,
\begin{equation}
T = \frac{\hbar}{mc^2} \sqrt{N}\label{e23}
\end{equation}
which is just equation (\ref{e2}).\\
We have rederived equation (\ref{e2}) but from a slightly different point
of view. Earlier we had started with the random walk equation (\ref{e1})
and then using the relation $R = cT,$ we got equation (\ref{e2}). In the
case of (\ref{e23}) we started with the well known fact that the fluctuation
in particle number is $\sim \sqrt{N}$. Infact as $N$ itself can be arbitrary
in principle, this shows that the two apparently different formulations are the same.
This will be further confirmed below.\\
We now observe that from (\ref{e22}) we get,
$$\dot R = \frac{G\dot Nm}{c^2} + \frac{Nm}{c^2} \dot G$$
As will be seen (cf. equation (\ref{e30})), $|\dot G|\sim N^{-1}$
while $\dot N$ from above is $ \sim \sqrt{N}$ and the second term is half the
first and with opposite sign. So
\begin{equation}
\frac{dR}{dt} \approx  HR\label{e24}
\end{equation}
and, from (\ref{e24}),
\begin{equation}
\frac{d^2R}{dt^2} \approx H^2R + \dot H R  = \lambda R, \lambda
\approx 0\label{e25}
\end{equation}
where $H$ in (\ref{e24}) can be identified with the Hubble Constant, and as
can be seen from the above (equations (\ref{e22}) ff.),
\begin{equation}
H = \frac{Gm^3c}{2 \hbar^2}\label{e26}
\end{equation}
Both the terms in (\ref{e25}) are equal and opposite and cancel out, that is
$\lambda$ vanishes.\\
Comparing (\ref{e25}) with the well known equation of General Relativistic
Cosmology\cite{r25}, we can see that $\lambda$ plays the role of the
cosmological constant which vanishes. However, in view of the order
of magnitude calculations, strictly speaking,
\begin{equation}
\lambda \le 0 (H^2)\label{e27}
\end{equation}
Indeed (\ref{e27}) agrees with observation\cite{r19}.\\
We emphasize that equations (\ref{e22}) and (\ref{e23}) show that in this formulation, the correct
radius and age of the universe can be deduced given $N$ as the sole
cosmological or large scale parameter.\\
Further, equation (\ref{e26}) or equivalently,
$$m \approx (\frac{\hbar^2 H}{Gc})^{1/3},$$
which is considered to be a mysterious and as yet unexplained coincidence is
rather a consequence. We will comment on this shortly.\\
To proceed we observe that the fluctuation of $\sim \sqrt{N}$ (due to the ZPF)
leads to an extra electrostatic energy $\frac{e^2 \sqrt{N}}{R}$ which is
balanced by the energy of the electron itself viz., $m_ec^2$\cite{r32}. Whence
we get:
\begin{equation}
R = \sqrt{N}l\label{e28}
\end{equation}
and
\begin{equation}
N_1 = \sqrt{N}  = \frac{e^2}{Gm^2} \approx 10^{40}\label{e29}
\end{equation}
which are just (\ref{e1}) and (\ref{e17}).\\
We have thus rederived the random walk equation (\ref{e1}) also, using
fluctuation in particle number, and in the process rederived (\ref{e17})
also.\\
If we combine (\ref{e28}) and (\ref{e22}), we get,
\begin{equation}
\frac{Gm}{lc^2} = \frac{1}{\sqrt{N}}\label{e30}
\end{equation}
So, as noted in the derivation of (\ref{e24}), $|\dot G| \sim N^{-1}.$
If we combine (\ref{e30}) and (\ref{e23}), we get Dirac's equation\cite{r14}
\begin{equation}
G \alpha T^{-1}\label{e31}
\end{equation}
It must be mentioned that, as argued by Dirac and Melnikov\cite{r37}
we treat $G$ as the variable, rather than the quantities $m (\mbox{or} l),
c \mbox{and}\quad \hbar$ (the micro physical constants), because of their central role
in atomic (and sub atomic) physics.\\
Further, using (\ref{e30}) in (\ref{e29}), with $N_1 = \sqrt{N}$,
we can see that the charge $e$ also is independant
of time or $N$. So $e$ also must be added to the list of microphysical
constants.\\
Next if we use $G$ from (\ref{e30}) in (\ref{e26}), we can see that
\begin{equation}
H = \frac{c}{l} \quad \frac{1}{\sqrt{N}}\label{e32}
\end{equation}
Thus apart from the fact that $H$ has the same inverse time dependance on
$T$ as $G$, (\ref{e32}) shows that given the microphysical constants, and
$N$, we can deduce the Hubble Constant also.\\
Using (\ref{e2}) and (\ref{e28}), we can deduce that
\begin{equation}
\rho \approx \frac{m}{l^3} \quad \frac{1}{\sqrt{N}} \propto T^{-1}\label{e33}
\end{equation}
Thus we have the scenario of an ever expanding (and possibly accelerating)
universe\cite{r38} with, as can be seen from (\ref{e33}), decreasing density.
Latest observations of distant supernovae by different teams of observers
confirm all this\cite{r39,r40}.\\
It should also be noted that in the above considerations we have deduced
theoretically the so called Dirac large number of coincidences, which have
been supposed to be accidental features. However the resulting cosmology
of Dirac was, as is well known\cite{r25} inconsistent owing to the fact
that there were relations like, $R \propto T^{1/3}$, which ofcourse is not
correct. But then it must be remembered that Dirac had considered a fixed
number of particles.
\section{Discussion}
1. It was already pointed out in section 2 that if the minimum cut off tends
to zero, we recover the conventional picture. In particular,
we can recover the usual Big Bang formulation in the limit $\tau \to 0$.
Indeed in this case we can see from the equation leading to (\ref{e23})
that $\frac{dN}{dt} \to \infty,$ indicating a singular creation. More
accurately, we take the Planck scale $(l_P, \tau_P)$ instead of the pion
scale $(l, \tau)$ as above
as the zero point scale as in quantum gravity\cite{r41}.
The Planck density $\rho_P$ is given by another mysterious "coincidence",
\begin{equation}
\rho_P \times l^3 = M\label{e34}
\end{equation}
It would also follow that the number of Planck masses in the above volume
$l^3$ is $N' \sim 10^{60}$ but then we have to consider the fact that in
the physical time period $\tau$ corresponding to the length $l$, there are
$\frac{\tau}{\tau_P} \sim 10^{20}$ Planck life times, so that the number
of particles in the physical interval $(l, \tau)$ is $N \sim 10^{80}$,
which is the total particle number. In the above derivation, we consider
the semi classical Quantum Gravity picture in conjunction with the
Quantum Mechanical Compton scale.\\
Thus we recover the entire mass and also the entire number of particles
in the universe, with a singular creation, as in the Big Bang theory. However
in this limit we cannot deduce the large number coincidences or the Weinberg
formula (\ref{e26}).\\
2. We have seen the intimate relationship between the Brownian type relations
(\ref{e1}) and (\ref{e2}), as also the fluctuation in particle number
$\sim \sqrt{N}$ on the one hand and purely quantum mechanical micro physics
as also cosmological considerations on the other. This is typified by not
just equations (\ref{e1}) and (\ref{e2}), but also equations like (\ref{e26}),
(\ref{e29}) and (\ref{e30}). This is an expression of what has been called
the micro-macro nexus\cite{r14} or stochastic holism, in contrast to
the conventional idea of a whole built out of parts as for example could be
typified by an equation like (\ref{e3}).\\
3. One may wonder how the cosmic microwave background radiation comes in.
Infact it has been shown that the observed cosmic microwave background
radiation is consistent in the above scheme\cite{r42,r14, r43}.\\
4. An important question that arises is that of the observed mass spectrum.
It has been shown, in a phenomenological scheme consistent with decay modes
that such a mass spectrum can indeed be consistently obtained\cite{r44}
in the light of foregoing considerations.\\
5. It is interesting to note that given the random walk relation (\ref{e1}),
using the relation (\ref{e23}) (rather than (\ref{e2})), we can deduce that
$R = cT$. In other words the velocity of light itself is seen to emerge
from the above stochastic formulation. Indeed this is not surprising because
as noted in section 2, such a maximum velocity is a consequence of discrete
space time.\\
6. In section 3 we obtained equation (\ref{e13}) (or (\ref{e14})) which was
the linearized equation of General Relativity. As is known starting from
such a linearized theory we could bootstrap our way to the full equation of
General Relativity\cite{r19,r45}.\\
7. In effect the minimum space time intervals have fudged the singularities
of General Relativity - for the Kerr-Newman formulation as also for the Big
Bang. Indeed as has been observed, these singularities of General Relativity
pose the greatest challenge of modern Physics. Such singularities are now
seen to be a result of going down to arbitrarily small space time intervals,
a process which is not legitimate even in Quantum Mechanics. At the same time,
the ultra violet divergences of Quantum Theory also disappear, which indeed
was one of the original motivations for space-time discretization (cf.refs.
\cite{r4,r9}).\\
8. We could derive the weak interactions from a slightly different point of
view. If these are mediated by an intermediate particle of mass $m_w$ and
Compton wavelength $l_w$ then, as in the run up to equation (\ref{e28}) and
(\ref{e29})
using the fluctuation in particle number we can deduce instead of (\ref{e29}),
now the equation\cite{r28}
\begin{equation}
g^2 \sqrt{N_\nu} l^2_w \approx m_\nu c^2\label{e35}
\end{equation}
where $N_\nu$ is the number of neutrinos and $m_\nu$ is the neutrino mass.\\
We use the fact that\cite{r46,r32}
$$m_\nu \approx 10^{-8} m_e,$$
that is the mass of the neutrino is about one hundred millionth that
of the electron, as indeed has been confirmed by the recent super Kamiokande
experiments. It is also known that $N_\nu \sim 10^{90}$ so that from
(\ref{e35}) we get,
$$g^2 l^2_w \sim 10^{-59}$$
which infact gives back equation (\ref{e21}) for the weak interaction.
\section{Conclusion}
Using the Brownian theory we have deduced minimum space time intervals
which have directly lead to the Dirac equation and a unified formulation of
not just quantum theory but quarks and leptons. In the process hitherto
purely empirical results like the discreteness of the charge, the handedness of the neutrino, or the
handedness of quarks, their fractional charge, confinement and masses have
been deduced from the theory. The so called accidental relations like the
classical Kerr-Newman metric providing a description of the field of the
electron including the purely quantum mechanical gyromagnetic ratio $g = 2$
as also the various large number coincidences and Weinberg's mysterious formula
are also now seen to be natural consequences of the theory.\\
This apart a cosmology consisting with observation is also seen to follow.

\end{document}